\documentclass{article}

\usepackage[final]{nips_2016} % produce camera-ready copy
\usepackage{floatrow}
\usepackage[utf8]{inputenc} % allow utf-8 input
\usepackage[T1]{fontenc}    % use 8-bit T1 fonts
\usepackage{hyperref}       % hyperlinks
\usepackage{url}            % simple URL typesetting
\usepackage{booktabs}       % professional-quality tables
\usepackage{amsfonts}       % blackboard math symbols
\usepackage{nicefrac}       % compact symbols for 1/2, etc.
\usepackage{microtype}      % microtypography
\usepackage{blindtext}
\usepackage{graphicx}
\usepackage{caption}
\usepackage{subcaption}
\usepackage{float}
\usepackage{xcolor}
\usepackage{multirow}

\title{Completing partial recipes using item-based collaborative filtering to recommend ingredients}

\author{
  Paula Ferm\'in Cueto\thanks{Equal Contribution}\\
  \texttt{paula.fermin.cueto@gmail.com} \\
  \And
  Meeke Roet$^{*}$\\
  \texttt{roet.meeke@gmail.com} \\
 \And
  Agnieszka Słowik$^{*}$\\
  \texttt{agnieszka.slowik@cl.cam.ac.uk} \\
}

\begin{document}

\maketitle

\begin{abstract}  
Increased public interest in healthy lifestyles has motivated the study of algorithms that encourage people to follow a healthy diet. Applying collaborative filtering to build recommendation systems in domains where only implicit feedback is available is also a rapidly growing research area. In this report we combine these two trends by developing a recommendation system to suggest ingredients that can be added to a partial recipe. We implement the item-based collaborative filtering algorithm using a high-dimensional, sparse dataset of recipes, which inherently contains only implicit feedback. We explore the effect of different similarity measures and dimensionality reduction on the quality of the recommendations, and find that our best method achieves a recall@10 of circa 40\%.
\end{abstract}

%-----------------------------------------------------------------------
% INSTRUCTIONS FROM TEMPLATE
%-----------------------------------------------------------------------

%The report should use this template and be 8 pages in length. Do not change the fontsize or layout. It should be compilable with pdflatex.

%Structuring the text as follows is likely useful, but definitely
%\emph{not} a requirement.

%-----------------------------------------------------------------------
% INTRODUCTION
%-----------------------------------------------------------------------

\section{Introduction}
\label{sec:intro}
The application of recommendation systems in the context of recipe and ingredient recommendation is becoming increasingly popular. With obesity being a major health problem across the globe and an increased public interest in healthy lifestyles, the study of algorithms that assist people in following a healthy diet is worthwhile. 

Most of the existing models that use recipe data aim to recommend recipes to users based on previous ratings of other users and recipes. For example, \cite{Freyne2010} developed recommender systems to suggest healthy recipes using neighbourhood-based collaborative filtering models. Although less popular in the literature, the task of predicting combinations of ingredients is also of interest. It is based on the idea that oftentimes, people have to make a shopping list whilst taking into account the products they still have at home. In such cases, the question is what additional ingredients could be bought to make a meal. 

There are websites, for example \href{https://www.allrecipes.co.uk}{\emph{allrecipes.co.uk}} and \href{https://www.epicurious.com}{\emph{epicurious.com}}, where one can input a list of ingredients and obtain recipes including those ingredients. However, this requires that a recipe exists in the database containing all the inputted ingredients, which may not be the case if the input is many ingredients. Hence, our goal is to build a recommendation system to suggest ingredients that form a good combination with a set of other ingredients. This recommendation system will treat recipes as \emph{users}, while the ingredients will play the role of the \emph{items} these recipes have `rated'.

One of the most popular methods for recommendation systems is Collaborative Filtering (CF), which uses a matrix of all items and users to produce recommendations, where typically each element of the matrix represents how much a user likes an item. Such methods are known to require a large volume of rating data to produce meaningful recommendations. We will use a recipe dataset provided by \emph{yummly} for the Kaggle competition \href{https://www.kaggle.com/c/whats-cooking}{`\emph{What's Cooking?}'}, which contains a large collection of recipes (circa 40k) and a list of ingredients included in each recipe. 

\cite{Su2009} classify CF algorithms into three categories: \emph{memory-based models}, which calculate the similarity between users and items from the entire rating matrix, \emph{model-based models}, which fit a parametric model to a given matrix of ratings to make predictions, and hybrid models. In the class of model-based algorithms, Matrix Factorisation (MF) is the most popular. MF characterises items and users by vectors of latent factors derived from patterns of item rating. This is the approach used for example by \cite{DeClercq2016} in a setting very similar to ours. They used a variety of MF approaches to predict combinations of ingredients for given partial recipes.

% such as Singular Value Decomposition (SVD), Non-negative Matrix Factorisation (NMF) and Independent Component Analysis (ICA)

In this project, we explore the other direction: memory-based collaborative filtering (MBCF), which has the advantages that it does not rely on item content, it is easy to implement and new data can be added incrementally, as described by \cite{Su2009}. Memory-based (or neighbourhood-based) models can be classified into two groups: \emph{item-based} approaches, which compute the similarity between items based on users' ratings of those items, and \emph{user-based} approaches, which focus on finding similar users based on the items that they rated, and use those to make predictions. Although user-based models were the first MBCF approach described, they do not scale well to large datasets and they are also not well suited for dealing with highly sparse data. To counter these issues, \cite{Sarwar2001} introduced the concept of item-based collaborative filtering and showed that this approach generally outperforms user-based models. Based on their evidence, we have selected the item-based approach as our preferred method for the task at hand.

%A positive characteristic of the recipe dataset is that the ratings of our `users' - recipes - are consistent and static over time, unlike the ratings of human users, whose tastes can be very diverse (e.g. ballet and heavy metal) and may change over time. The fact that our data is static could have a positive impact on the accuracy of our model.

One of the main challenges we face is the fact that our ratings are binary, positive-only (`implicit feedback'). Unlike other CF tasks, such as suggesting movies for users based on the ratings that other, similar users have given to other movies, the recipe dataset does not contain explicit user feedback. A given recipe has either rated an ingredient `positively' (value 1) or `has not seen it' (value 0), but we do not have negative ratings: a value of 0 for an ingredient in a recipe does not imply that the ingredient would not go well with the rest of the ingredients. The problem of dealing with such implicit feedback is becoming more and more relevant nowadays, because explicit feedback, although more convenient, is not always available, for example when dealing with click or purchase data. Fortunately, it is a well-studied issue. For instance, \cite{Aiolli2013} proposes an asymmetric cosine similarity measure that is well suited for this type of rating data.

%, \cite{Hu2008} propose a factor model specifically designed for positive-only data, proposing to treat this binary data as an indication of positive or negative feedback,

We have opted for experimenting with different types of similarity measures to deal with the particularities of our dataset. However, this is also interesting for more general reasons, as the computation of similarity between items or users is a critical component of MBCF techniques and has a big impact on the quality of the predictions. We will thus investigate the effects of using different similarity measures on the quality of the recommendations. Furthermore, we will explore the benefits of applying dimensionality reduction techniques to deal with sparse binary data.

%modifications to the standard item-based CF algorithm to tailor it for sparse binary data.

% To evaluate our model, we will hide one ingredient from a given recipe and will use our model to recover it. The model will return a list of suggested ingredients for that partial recipe, sorted by the degree to which they make a good combination with the remaining ingredients in the recipe. Our aim is that the hidden ingredients rank high in this list when we repeat this process for a number of recipes.

% description of the task/objective
% relevant background and related previous work
% explanation of the significance/relevance of the objective/task

%-----------------------------------------------------------------------
% DATA PREPARATION
%-----------------------------------------------------------------------

\section{Data preparation}
Our dataset is the training data from the `\emph{What's Cooking?}'  challenge on Kaggle. It contains 39,774 recipes identified by a recipe id and classified into 20 different cuisines. For each recipe, there is a list of its ingredients, and there are 6,782 distinct ingredients in total.

To apply collaborative filtering techniques on this dataset we need to convert it to a rating matrix $R$ of dimension $m \times n$, where $m$ is the number of recipes and $n$ is the number of ingredients. Element $r_{ui}$ will be a binary variable taking value 1 if ingredient $i$ is contained in recipe $u$ and 0 otherwise. This will thus be similar to the feature representation known as `bag-of-words', which is widely used in natural language processing. We used the \texttt{CountVectorizer} class from scikit-learn library to create this representation.

The second preprocessing step is to clean the data from irrelevant recipes and ingredients. Firstly, we removed 3,003 ingredients that appear three times or less, we cleaned the ingredient names by removing special characters and numbers (e.g. '14 oz') and corrected for the fact that some ingredients started with a space.

% , because they are so infrequent that they will only increase the computational burden of our approaches without adding significant information.

After these initial pre-processing steps, the list of ingredients was still noisy. The ingredient names included words that do not contain relevant information and only indicated different versions of the same thing. For example, `light brown sugar' and `dark brown sugar' should be combined into one ingredient `brown sugar', and words like `half' or `oz' should be removed from the name. We came up with the following approach to combine ingredients according to this intuition:

\begin{enumerate}
\item We create a list of smaller words contained in ingredient names by finding longest common substrings between all pairs of ingredients.
\item We simplify the name of each ingredient by keeping only the substrings that appear in more than 30 different ingredients or in more than 1,000 recipes. 
\end{enumerate}

By doing this we ensure that, for instance, `chicken-broth' will not change, since both `chicken' and `broth' are dominant (they appear often), but `adzuki beans' will be replaced with `beans', because `adzuki' is very rare. The thresholds were adjusted by qualitative evaluation of the results during trial and error. This process reduced the number of the ingredients by one-third, but the data was still very high-dimensional, and many ingredients were still relatively infrequent. 

To counter this issue, and given the large collection of recipes available to us, we removed all ingredients appearing in 250 recipes or less. We also removed ingredients that we considered similar to `stop words' in natural language processing: `salt' or `water' would not be very valuable recommendations for complementing a set of ingredients. Lastly, we removed all recipes that contained two ingredients or less. Recipes with no ingredients at all are errors and recipes with one or two ingredients do not allow for predictions and/or evaluation of these predictions. Our final dataset contains 37,340 recipes and 267 ingredients.

%-----------------------------------------------------------------------
% EXPLORATORY DATA ANALYSIS
%-----------------------------------------------------------------------

\section{Exploratory data analysis}
In order to understand the data we are using and to avoid wrong interpretation of future results we perform exploratory data analysis before moving to the collaborative filtering experiments. The data is binary, high-dimensional and sparse (on average, 0.03\% of the features in a recipe are non-zero), which makes descriptive statistics and visualisation particularly useful in extracting main characteristics of the population. 

% THIS NOW MAKES MORE SENSE AFTER THE T-SNE PLOT, SINCE WE DON'T HAVE THE PLOT WITH THE TYPES OF CUISINES HERE ANYMORE!

\begin{figure}[htb]    \centering
	\includegraphics[width=\textwidth]{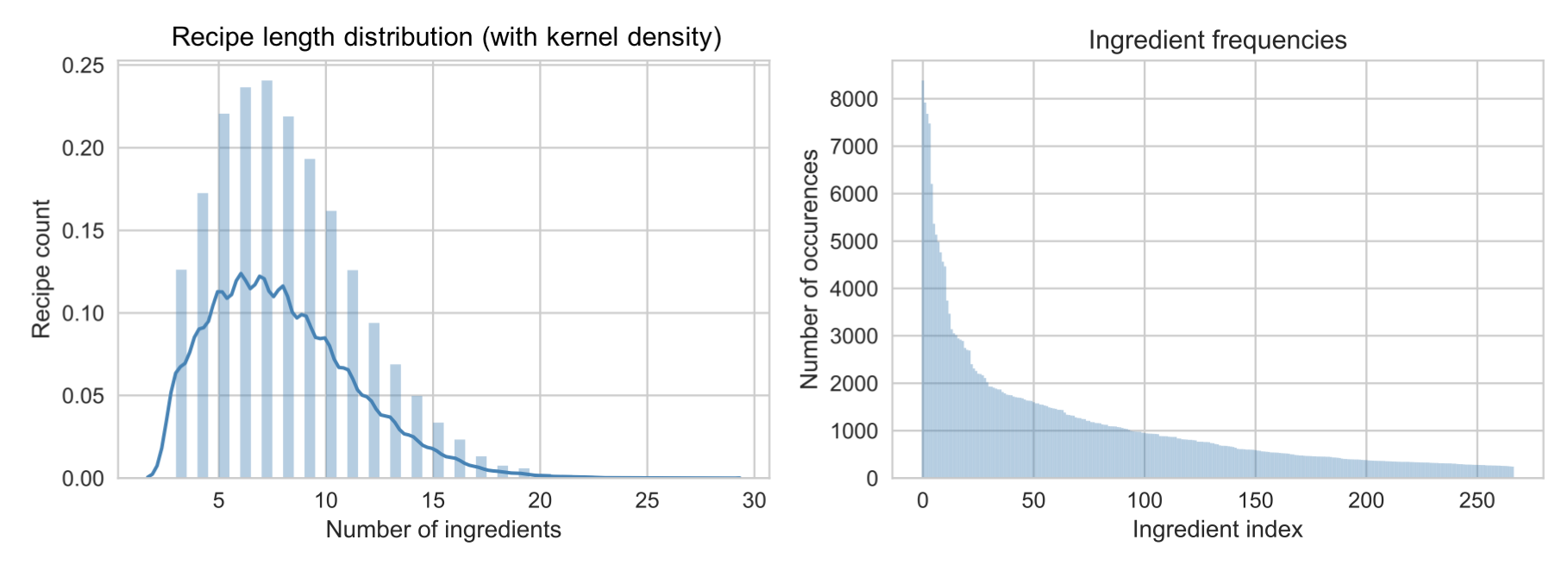}
    \caption{Data distribution.}
    \label{fig:ingredient_frequency}
\end{figure}

More than half of the ingredients are present in less than 1,000 recipes (Figure \ref{fig:ingredient_frequency}, right). There is still a small number of very popular ingredients that can be found in more than 8,000 recipes. Those ingredients are: \textit{butter, tomatoes, garlic, sugar, olive} and \textit{onions}. They are likely to be recommended most often by any model, since they seem to go well with more than 21.42\% of recipes in the training set. Those ingredients are also a typical base for an Italian dish, and Italian is the most popular cuisine in the training data.

The number of ingredients in a recipe (Figure \ref{fig:ingredient_frequency}, left) might also influence the recommendations. Both the median and the mean recipe length are 8.0 ingredients, while the minimum and maximum are 3 and 28, respectively. Recipes with less than 5 ingredients normally include a source of protein, and rice or bread, and they might not showcase all the subtle correlations that can arise when there is a higher range of ingredients to choose from. The recipe length distribution has moderate positive skewness of 0.81 and kurtosis of 0.76, suggesting that while the recipes tend to have a small number of ingredients, there are some outliers. Those recipes might introduce noise and bias to the results of similarity functions, especially to those that are based on the co-occurrences of ingredients: co-occurrence in a recipe of more than 20 ingredients in total might be less indicative of similarity between the ingredients.
% Removed PMI here, as we haven't introduced it yet.

%\begin{figure}[H]
%\centering
%  \begin{subfigure}{0.49\textwidth}
%    \includegraphics[width=\textwidth]{PCA.pdf}
%    \hspace*{\fill}
%    \label{fig:pca}
%  \end{subfigure}
%  \begin{subfigure}{0.49\textwidth}
%    \includegraphics[width=\textwidth]{t-sne_plot.png}
%    \hspace*{\fill}
%    \label{fig:tsne}
%  \end{subfigure}
%\hspace*{\fill}
%\caption{Data visualisation using dimensionality reduction.}
%\label{fig:vis}
%\end{figure}

\begin{figure}[htb]    \centering
	\includegraphics[width=\textwidth]{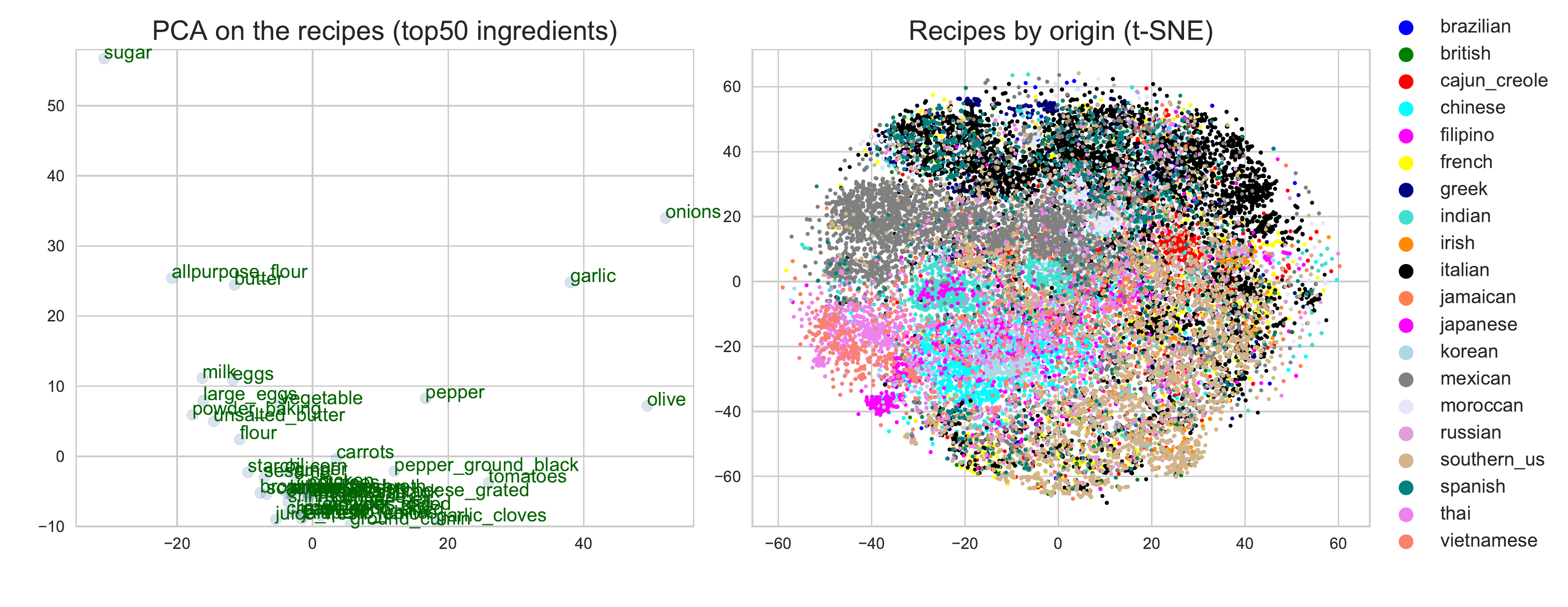}
    \caption{Data visualisation using dimensionality reduction.}
    \label{fig:vis}
\end{figure}

%To identify location and shape of similar recipes groups, we used dimensionality reduction algorithms to cluster recipes based on their ingredients.
% Location and shape - the notions from the lectures
%\ref{fig:pca}%
We used dimensionality reduction methods to visualise clusters of ingredients based on their occurrences in recipes. For this task, we used Principal Component Analysis on the recipes (Figure \ref{fig:vis}, left), implemented by using the \texttt{PCA} class in scikit-learn library, which carries out Singular Value Decomposition behind the scenes. For better interpretability, we display only the 50 most popular ingredients. The results make intuitive sense. For example, common baking ingredients -- \textit{milk, eggs, baking powder, unsalted butter, flour} -- appear in a distinct cluster. Moreover, the extremely common ingredients mentioned earlier were plotted far away from the main cluster.

%and found that the first 5 mapped to a two-dimensional plot produce the most interpretable visualization (while explaining $0.14\%$ of variance). 

To analyse the similarity between recipes of the same cuisine we tried different techniques to visualise very high-dimensional data in a 2-dimensional representation. First, we implemented PCA to reduce the number of ingredients, but this technique was not very successful. In the most interpretable visualisation we were able to observe that Asian and European were grouped together, but distinct cuisines were not separated. We then experimented with t-distributed Stochastic Neighbor Embedding (t-SNE), a visualisation technique that implements a heavy-tailed distribution in a space of reduced dimensionality, which has been demonstrated to outperform other non-parametric methods in multiple domains with high-dimensional data \citep{VanDerMaaten2008}.
%such as PCA, Isomap or Sammon Mapping 

As expected, t-SNE was a more suitable technique for representing our data and it can be seen in Figure \ref{fig:vis} (right) that it creates distinct clusters of most cuisines. This representation also captured subtler characteristics of cuisines -- for instance, the Indian cuisine is close to the rest of the Asian cuisines (Thai, Vietnamese, Japanese, Korean Filipino, Chinese), but it forms a separate cluster, since there are some ingredients that are very frequent in Indian food but not so common in Western Asian recipes. Similarly, Mediterranean cuisines are grouped in adjacent clusters with Moroccan recipes being clearly separated. We find the biggest overlap in Spanish and Italian recipes, which makes sense, since both make abundant use of olive oil, tomatoes, seafood and certain spices. Most of the recipes come from the Italian (20\%), Mexican (16\%) and Southern American (11\%) cuisines. The recommendations are likely to be biased towards the ingredients used in those cuisines.

%-- while they are influenced by the Spanish and French cuisine, they contain ingredients typical to Arab countries that are much less popular in Spanish, Italian and Greek recipes. IT IS TRUE BUT IT'S PROBABLY A BIT TOO SPECIFIC AND WE NEED TO MAKE SPACE.

%-----------------------------------------------------------------------
% LEARNING METHODS
%-----------------------------------------------------------------------

\section{Methodology}
\subsection{Item-based collaborative filtering}

%In our application, the recipes act as `users' and the ingredients are seen as `items'. This means 

%As discussed in the introduction the user-item matrix $R$ has dimension $m$ x $n$, where $m$ is the number of recipes and $n$ is the number of ingredients. Each entry $r_{ui}$ is a binary variable taking the value 1 if recipe $u$ contains ingredient $i$ and 0 otherwise.
%Section \ref{sec:number_of_neighbours} provides details on the choice of $k$.
% Section \ref{sec:similarity_measures} describes the similarity measures that we experiment with. 

% In this setting, these are binary vectors of length $m$ indicating whether or not an ingredient is present in a recipe.
Item-based CF is implemented in two steps. First, the similarity between all ingredients is calculated: for two ingredients $i$ and $j$, we calculate the similarity between the two vectors containing their `ratings'. Then, for each ingredient, the model selects the $k$ most similar ingredients, where the number of neighbours $k$ is a parameter defined by the user. In the second step, the fit $P(u,i)$ between ingredients $i$ and recipe $u$ is predicted (\citep{Sarwar2001}).

%The fit $P(u,i)$ of ingredient $i$ to recipe $u$ is calculated as \citep{Sarwar2001}:
\begin{equation}
  P(u,i) = \frac{\sum_{j \in N_i} s_{ij} \cdot r_{uj}}{\sum_{j \in N_i} |s_{ij}|}
\end{equation}
where $N_i$ is the set containing the $k$ neighbours most similar to $i$, and $s_{ij}$ is the similarity between ingredient $i$ and $j$. The fit of an ingredient to a recipe is thus obtained by summing the similarities of the target ingredient's closest neighbours that are present in the target recipe and then scaling by the sum of all similarities in the target ingredient's neighbourhood to obtain values between 0 and 1. The fit is 0 if none of the ingredient's closest neighbours appear in a recipe, and 1 if all of its neighbours appear. The more close neighbours of an ingredient appear in a recipe, the higher the predicted fit of the ingredient to that recipe. Closer neighbours will contribute to the fit more, as their higher similarity increases the numerator more. Finally, recommendations for a recipe are made by selecting the $N$ ingredients with the highest predicted fit, with $N$ the desired number of recommendations.

\subsection{Similarity measures}
\label{sec:similarity_measures}

One part of our experiment is exploring the effect of different similarity measures on the performance of item-based CF. The first measure was suggested by \cite{Sarwar2001} at the introduction of item-based CF. The \textbf{cosine similarity (CS)} between two ingredients $i$ and $j$ is:
\begin{equation}
  CS_{ij} = \frac{r_{:i} \cdot r_{:j}}{||r_{:i}||_2 \cdot ||r_{:j}||_2}
   		 = \frac{|U(i) \cap U(j)|}{|U(i)|^{1/2} |U(j)|^{1/2}} 
\end{equation}
where $r_{:i}$ is the column vector containing all entries for ingredient $i$, taken from the user-item matrix $R$, `$\cdot$' indicates the dot product, and $U(i)$ is the set of recipes containing ingredient $i$. Because all data is binary, the expression simplifies to the number of recipes containing both ingredient $i$ and $j$ divided by the square root of the number of recipes containing $i$ times the number of recipes containing $j$ \citep{Aiolli2013}. The cosine similarity is symmetric and efficient to calculate in sparse spaces. The second characteristic is positive for our application, but the first might be problematic \citep{Aiolli2013}. For example, if a recipe contains a very uncommon ingredient, then common ingredients that are similar will probably go well with the recipe. The opposite, however, does not necessarily hold. % and the uncommon ingredient should probably be recommended less often. 

For this reason, \cite{Aiolli2013} proposes the \textbf{asymmetric cosine similarity (ACS)}:
\begin{equation}
  ACS_{ij} = \frac{|U(i) \cap U(j)|}{|U(i)|^{\alpha} |U(j)|^{1-\alpha}}
\end{equation}
where $\alpha$ is the asymmetry parameter, which puts more weight on common ingredients as it increases. It is expected that tuning $\alpha$ will provide an advantage in our dataset, because the least common ingredient occurs 250 times and the most common ingredient over 8,384 times (after data preparation). The tuning is done by carrying out a parameter sweep over values from 0 to 0.5, as suggested by the author.% on 10\% of the dataset.

The third similarity measure we investigate is \textbf{Jaccard similarity (JS)}:
\begin{equation}
  JS_{ij} = \frac{|U(i) \cap U(j)|}{|U(i) \cup U(j)|}
\end{equation}
The main downside of this measure is that it loses information when the magnitude of ratings is meaningful, because it only counts what has and has not been rated to determine similarity. In our case, this is not a problem, as similarity between ingredients is loosely defined as appearing in the same recipes `often', which is precisely what Jaccard similarity measures.

Lastly, \cite{Bellogin2011} showed that measures from information theory are also suitable for collaborative filtering. To add variety to the set of similarity measures that is experimented with, we add the \textbf{pointwise mutual information (PMI)} to the selection:
\begin{equation}
  PMI_{ij} = log \frac{p(i,j)}{p(i)p(j)}
\end{equation}
where $p(i,j)$ is the probability of co-occurrence of ingredient $i$ and $j$ in a recipe, and $p(i)$ is the probability of ingredient $i$ being present. The PMI measures how often ingredients are used together relative to their frequency of usage in general. It is therefore a very intuitive measure for our purpose. One downside is that it is symmetric, which could reduce its performance as explained earlier.

\subsection{Principal Component Analysis}
\label{sec:PCA}

\cite{Candillier1997} describes how dimensionality reduction techniques such as Principal Component Analysis (PCA) can be applied to neighbourhood-based CF to obtain results of better quality with less computational effort. The reason behind this is that using PCA to reduce the dimensionality of the data speeds up the computation of the similarity matrix and makes these computations more robust, as it enables the transformation of a highly sparse representation to a dense, low-dimensional model of latent factors. Given the high dimensionality of our ratings data, especially the number of recipes, we expect that both these benefits apply in our case. Therefore, the second part of our experiment is to apply the methodology outlined in the previous sections both on the dataset immediately after preprocessing, and after applying PCA to the rows (recipes) and compare the results.

Once again we use the \texttt{PCA} class in scikit-learn library to obtain a transformed representation of the recipe space consisting of 267 principal components, as many as there are ingredients. This cannot be more, because the maximum number of principal components is the rank of a matrix, in our case the number of ingredients.

\subsection{Number of neighbours}
\label{sec:number_of_neighbours}

The number of neighbours that is used for predicting the fit of ingredients to recipes can have a great influence on the performance of CF. If $k$ is too low, the predictions will have a high variance and be biased towards only a few instances. If $k$ is too high, the predicted fits will all be very similar, so the ranking will be susceptible to small changes. Therefore, we optimize $k$ separately for each similarity measure and original/PCA-reduced data by performing a parameter sweep over the set $\{10, 20, 50, 100, 150, 200\}$ using 10\% of the dataset. This should suffice, as we do not need to `train' our model in a classical sense to obtain accurate parameters, but only need a ranking of the performance under different values of $k$.

\subsection{Evaluation}
\label{sec:evaluation}

The quality of the recommendations under different parameters, similarity measures, and original/PCA-reduced data is evaluated by means of a leave-one-out cross validation scheme. We select one recipe at a time, which we take out of the dataset. Then, we remove one ingredient at random from the recipe in question and make recommendations for the partial recipe based on the remaining dataset. Of interest is how good the recommendation algorithm is at recovering the missing ingredient. Following \cite{DeClercq2016}, this is assessed using three metrics:
\begin{enumerate}
  \item Percentage of recipes for which the missing ingredient is among the top-10 recommended ingredients, also called recall@10 (the logic being that in a practical application of our work, ten ingredients would be a reasonable number of recommendations to make to a user looking for suggestions).
  \item Mean rank of the missing ingredients in the list of recommended ingredients.
  \item Median rank of the missing ingredients in the list of recommended ingredients.
\end{enumerate}

The first metric should be as high as possible, the other two as low as possible. After experimenting with different parameter settings on 10\% of the data, we select the most promising configurations and evaluate their performance using the other 90\% of the dataset as a test set. In addition to the quantitative metrics described above, we will also sample a few recipes and judge the corresponding recommendations qualitatively, simulating real-world use of our system.

%-----------------------------------------------------------------------
% RESULTS
%-----------------------------------------------------------------------

\section{Results}
\begin{figure}[htb]    \centering
	\includegraphics[width=\textwidth]{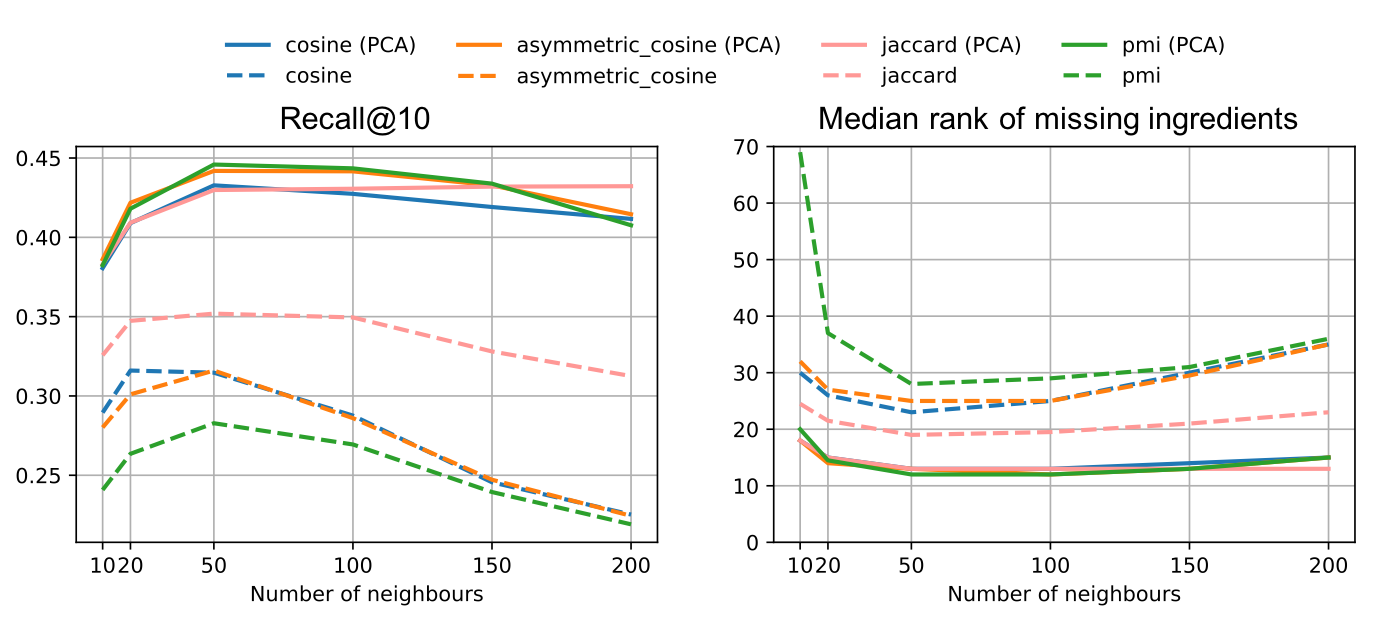}
    \caption{Effect of number of neighbours and PCA on recall@10 and median rank performance.}
    \label{fig:k_tuning}
\end{figure}

%with on the left the percentage of missing ingredients in the top 10 and on the right the median rank of the missing ingredient 
Experimenting with the parameter $\alpha$ for the asymmetric cosine similarity showed that the best performance was obtained when setting $\alpha = 0.05$. Figure \ref{fig:k_tuning} plots the results of tuning $k$ for all similarity measures and the original/PCA-reduced data. We decided to show the median rank here as it is more robust against outliers, but the mean rank showed similar patterns.

% It can be seen that the asymmetric cosine does not outperform the cosine as we expected. This might explain why the measure is not widely used in the literature

%so for example the distinction between the $10$th and the $11$th closest neighbour becomes less clear

%This makes sense, because with highly sparse binary data, the similarities between ingredients are overconfident, but with a low rank dense representation the ratings become smooth and adding additional neighbours has subtler effect.

%which makes sense, because by reducing the number of recipes, PCA makes it harder to distinguish between ingredients. For example the distinction between the $10$th and the $11$th closest neighbour becomes less clear.

It can be seen that the approaches using the PCA-reduced data clearly and consistently outperform those using the non-reduced data. Hence, we continue the analysis using only the PCA-reduced data. It can also be seen that, after applying dimensionality reduction, the asymmetric cosine outperforms the cosine as we expected. Another observation is that the algorithm becomes less sensitive to the number of neighbours after PCA. This makes sense, because with sparse binary data, the similarities between ingredients are overconfident, but in a dense representation the ratings are smooth and additional neighbours have a subtler impact. Also, the experiments with the non-reduced data took 141 minutes to run, whereas those using the PCA-reduced data took only 12 minutes. PCA is thus a substantial improvement in speed, quality, and robustness. We select $k = 50$ for all similarity measures, since 50 makes a slightly better impression than 100 in the plots and it is more sensible in proportion to the total number of ingredients.

Table \ref{similarity selection} shows the results of all similarity measures with the PCA-reduced test data using $k = 50$. The metrics are all close together. Based on the recall@10 and median rank, PMI and asymmetric cosine appear to perform best, and the cosine and Jaccard similarities perform slightly worse. Interestingly enough, the mean rank of PMI is the worst of the four, so it may have had a few more cases where it was clearly off, whereas the simple cosine does best on this. Depending on the specific application, any of the similarity measures could be useful. In our case, we would pick PMI, putting most weight on the robust metrics recall@10 and median rank.

\begin{table}[htb]
\centering
\caption{Evaluation of similarity measures with $k = 50$ and PCA-reduced test data.}
\label{similarity selection}
\begin{tabular}{l r r r}
\hline
Similarity measure & Recall@10       & Mean rank & Median rank \\ \hline 
Cosine             & 39.26\%         & 40.91     & 16          \\ 
Asymmetric cosine  & 40.16\%         & 41.69     & 15          \\ 
Jaccard            & 39.26\%         & 41.22     & 16          \\ 
PMI                & 40.25\%         & 42.73     & 15          \\ \hline
\end{tabular}
\end{table}

The work most similar to ours is \cite{DeClercq2016}, who obtained a recall@10 of 57.5\% with two-step recursive least squares (RLS), and 43.6\% with non-negative matrix factorisation (NNMF) on a dataset with more ingredients, but also substantially more recipes. The higher score of RLS can be explained by the fact that it incorporated additional information about the ingredients that we did not have available, such as flavour and texture. NNMF did not do this, and still performs slightly better than our best model. We attribute this to a difference in cleanliness of the datasets. For example, even after preprocessing, our dataset still contained some duplicate ingredients, such as `tortilla' and `corn tortilla'. We expect that solving such issues would improve the results further and close the gap to \citeauthor{DeClercq2016}'s NNMF approach.

Qualitative evaluation of our results strengthens this belief. Table \ref{two-recommendations} shows two random recipes from our dataset and the corresponding recommendations produced by our best model. The `corn versus corn tortilla' issue shows for example with the cilantro, chicken and garlic ingredients. The algorithm could be improved by filtering such ingredients from the list of recommendations. Apart from this, however, the table paints a positive picture, because the recommendations make intuitive sense. It is interesting to see that although we did not provide the model with the recipe's cuisine, it seems to implicitly detect this and recommend related ingredients (avocado, lime, sour cream, corn and salsa are all very typical Mexican ingredients; chillies, chicken, potatoes and cumin are typical for an Indian curry).

\begin{table}[htb]
\centering
\caption{Samples of recommendations using PMI similarity and PCA-reduced data with $k = 50$.}
\label{two-recommendations}
\begin{tabular}{l|ll}
\hline
\multirow{2}{*}{\begin{tabular}[c]{@{}l@{}}Recipe 16903 \\ (Mexican) \end{tabular}   } & \begin{tabular}[c]{@{}l@{}}Current \\ ingredients\end{tabular}      & \begin{tabular}[c]{@{}l@{}}cheddar\_cheese,  jalapeno\_chilies, lettuce,  lime, pork,  \\ purple\_onion, peppers , olive, cilantro\_chopped\_fresh,\\ pepper\_ground\_black tortillas\_corn \end{tabular}                            \\ \cline{2-3} 
                                        & \begin{tabular}[c]{@{}l@{}}\textit{Recommended} \\ \textit{combinations}\end{tabular} & \textit{\begin{tabular}[c]{@{}l@{}}avocado, cilantro, tomatoes, juice\_fresh\_lime, kosher\_salt, \\ cream\_sour, cilantro\_fresh, corn, salsa, ground\_cumin \end{tabular}}                                         \\ \hline
\multirow{2}{*}{\begin{tabular}[c]{@{}l@{}}Recipe 13162 \\ (Indian) \end{tabular} }  & \begin{tabular}[c]{@{}l@{}}Current \\ ingredients\end{tabular}      & \begin{tabular}[c]{@{}l@{}}butter, cayenne\_pepper, cream, garlic\_paste, ground\_cumin, \\ masala, milk, oil, onions, shallots, pepper\_black, \\ skinless\_boneless\_chicken, powder\_chili, yogurt\end{tabular} \\ \cline{2-3} 
                                        & \begin{tabular}[c]{@{}l@{}}\textit{Recommended} \\ \textit{combinations}\end{tabular} & \textit{\begin{tabular}[c]{@{}l@{}}paprika, chilies\_green, coriander\_ground, chicken, flour, \\ potatoes, cilantro\_leaves, pepper, garlic, cumin\_seed\end{tabular}}                                          \\ \hline
\end{tabular}
\end{table}

%-----------------------------------------------------------------------
% CONCLUSIONS
%-----------------------------------------------------------------------

%\section{Conclusions}
\section{Conclusions}
This project investigated the use of item-based collaborative filtering to recommend ingredients to partial recipes using sparse, high-dimensional data. We implemented four different similarity measures and used our algorithm directly on the preprocessed data as well as after an additional dimensionality reduction step. We found that our algorithms produce very intuitive recommendations. Reducing the number of recipes through PCA improves the quality of the recommendations substantially, while cutting the runtime by a factor 10 and making the results more robust against changes in the number of neighbours and the similarity measure. The best similarity measure depends on the requirements of a specific application. Our best method, using similarity measure PMI with $k=50$ and PCA, achieves a recall@10 of 40\%. This is slightly worse than the non-negative matrix factorisation approach taken by \citet{DeClercq2016}, but we believe that this gap can be closed, and potentially turned around to our advantage, by improving the quality of the dataset.

\bibliography{Edinburgh-DME}

\newpage

%\section*{Contribution of team members}

%\begin{itemize}
%\item \textbf{Agnieszka Slowik}:
%	\begin{itemize}
%	\item Report: Exploratory data analysis.
%    \item Code: Exploratory data analysis, evaluation metrics, exploration of alternative methods that did not make the cut.
%	\end{itemize}
%\item \textbf{Meeke Roet}:
	%\begin{itemize}
	%\item Report: Methodology, results (text).
    %\item Code: Similarity measures, collaborative filtering algorithm, support with data preparation.
	%\end{itemize}
%\item \textbf{Paula Fermin Cueto}: 
	%\begin{itemize}
	%\item Report: Introduction, data preparation, results (plots and tables).
    %\item Code: Data preparation, experiments + processing, support with similarity measures.
	%\end{itemize}

%\end{itemize}
%Literature review, code validation, editing and proofreading were carried out by everyone.

\end{document}